\documentclass{PoS}

\newcommand{\lambdar}{\lambda_{\mathrm R}}
\newcommand{\mr}{m_{\mathrm R}}

\title{Displacement Operator Formalism}

\ShortTitle{Displacement Operator Formalism}

\author{\speaker{Joannis Papavassiliou}\\
         Departamento de F\'\i sica Te\`orica and IFIC Centro Mixto,
        Universidad de Valencia--CSIC, E-46100, Burjassot, Valencia, 
	Spain\\
        E-mail: \email{Joannis.Papavassiliou@uv.es}}

\author{Daniele Binosi\\
       ECT*, Villa Tambosi, Strada delle Tabarelle 286
       I-38050 Villazzano (Trento), Italy \\ 
       I.N.F.N., Gruppo Collegato di Trento, Trento, Italy \\
        E-mail: \email{binosi@ect.it}}

\author{Apostolos Pilaftsis\\
CERN, Physics Department, Theory Division, CH-1211 Geneva 23, Switzerland,\\ 
 School of Physics and Astronomy, University of
Manchester,\\ Manchester M13 9PL, United Kingdom  \\      
         E-mail: \email{Apostolos.Pilaftsis@manchester.ac.uk}}

\abstract{ We review the $D$-formalism, 
a new method for  determining the renormalization of Green
functions to   all orders in  perturbation  theory.
This formalism exploits the fact  that the renormalized Green functions may
be calculated  by  displacing by  an infinite amount  the renormalized
fields and parameters of the theory with respect to the unrenormalized
ones. We briefly discuss some of its main characteristics, its conceptual and
practical advantages, and some new results obtained through it.}

\FullConference{International Europhysics Conference on High Energy Physics\\
		 July 21st - 27th 2005\\
		 Lisboa, Portugal}

\begin{document}

We review a new  formalism, the {\em
Displacement Operator Formalism}, or  the $D$-formalism, 
which 
enables one to  systematically organize  and explicitly
compute  the  counterterms (CT)
involved  in  the  renormalization
procedure, to  {\it all orders}  in perturbation theory.   The central
observation, leading  to this new  formulation, is that the  effect of
renormalizing any  given Green function  may be expressed in  terms of
ultraviolet infinite  displacements caused by the renormalization
on  {\it  both}   the  fields  and  the  parameters   of  the  theory.
Specifically, these UV infinite displacements, or shifts, quantify the
difference   between   fields   and   parameters  before   and   after
renormalization. We will present 
the derivation of the D-formalism in the context of a 
$\lambda\phi^4$ theory;  for the generalization  to  more involved
field theories, see~\cite{Binosi:2005yk}.

Let  $\Gamma_{\phi^n}$ be a bare,  one-particle irreducible 
$n$-point  Green  function.  After  carrying  out the  renormalization
programme, in $d=4-\epsilon$ dimensions, one will have that
\begin{equation}
  \label{Gren}
\phi^n\Gamma_{\phi^n}(\lambda,m^2;\mu,\epsilon)\ =\ \phi^n_{\mathrm{R}}
\Gamma_{\phi^n}^{\mathrm{R}}(\lambdar,\mr^2;\mu)\; , 
\end{equation}
where   $\lambda,  m^2,     \phi$  (respectively  $\lambdar,    \mr^2,
\phi_{\mathrm{R}}$) are the bare (respectively renormalized) parameters
and dynamical field   of the theory   at hand,
\begin{equation}
\phi=Z_\phi^{\frac12}\phi_{\mathrm{R}}, \qquad
\lambda=Z_\lambda\lambdar, \qquad m^2 = Z_{m^2}\mr^2, 
\label{CTs}
\end{equation} 
and    $\Gamma_{\phi^n}^{\mathrm{R}}(\lambdar;\mu)$   represents   the
renormalized $n$-legs Green function.  
Then,  we may  rewrite  the  left-hand  side of
(\ref{Gren}) as
\begin{equation} 
\phi^n\Gamma_{\phi^n}(\lambda,m^2;\mu,\epsilon)\ =\
(\phi_{\mathrm{R}}+\delta\phi)^n
\Gamma_{\phi^n}(\lambdar+\delta\lambda,\mr^2 + \delta m^2;\mu,\epsilon)\;, 
\end{equation} 
where the parameter shifts 
\begin{equation}
\delta\phi=(Z_\phi^{\frac12}-1)\phi_{\mathrm{R}}, \qquad
\delta\lambda=(Z_\lambda-1)\lambdar, \qquad \delta
m^2=(Z_{m^2}-1)\mr^2, 
\label{shmore}
\end{equation} 
quantify the difference of the renormalized quantities with respect to
the  corresponding  unrenormalized  ones  in a  given  renormalization
scheme R. Notice that our formulation treats fields
and couplings on equal footing, {\it i.e.}, as independent fundamental
parameters of the theory.

Next we use a Taylor expansion  to trade the combinations
$\phi_{\mathrm{R}}+\delta\phi$,      $\lambdar+\delta\lambda$,     and
$\mr+\delta m$,  for the renormalized  parameters $\phi_{\mathrm{R}}$,
$\lambdar$, and  $\mr$.  To this end,  it is natural  to introduce the
differential displacement operator
\begin{equation}
\label{Ddef}
D\ =\ \delta\phi\frac\partial{\partial\phi_{\mathrm{R}}}+
\delta\lambda\frac\partial{\partial\lambdar}+\delta
m^2\frac{\partial}{\partial \mr^2}\; ,
\end{equation}
in which the shifts are treated as independent parameters, and only at
the very end of all the manipulations they will be replaced by their
actual values in terms of the renormalization constants
$Z_\phi(\lambdar,\mr^2;\epsilon)$,
$Z_\lambda(\lambdar,\mr^2;\epsilon)$, 
and $Z_m(\lambdar,\mr^2;\epsilon)$.  It is then not difficult to derive
the following (all-order) master equation
\begin{equation}
  \label{master}
(\phi_{\mathrm{R}}+\delta\phi)^n\,\Gamma_{\phi^n}(\lambdar +
\delta\lambda,\mr^2+\delta m^2;\mu,\epsilon) \ =\
\Big<\,e^D\,\phi^n_{\mathrm{R}}\,
\Gamma_{\phi^n}(\lambdar,\mr^2;\mu,\epsilon)\, \Big>\;,
\end{equation}
or equivalently            
\begin{equation}
  \label{master2}
\phi^n_{\mathrm{R}}\,
\Gamma_{\phi^n}^{\mathrm{R}}(\lambdar,\mr^2;\mu)\ =\ 
\Big<\,e^D\,\phi^n_{\mathrm{R}}\,
\Gamma_{\phi^n}(\lambdar,\mr^2;\mu,\epsilon)\, \Big>\;, 
\end{equation}
where   $\langle  \dots   \rangle$  means   that  the   CT  parameters
$\delta\phi$,  $\delta \lambda$,  and $\delta  m^2$ are  to be  set to
their actual  values according to  (\ref{shmore}) after the  action of
the $D$  operator. 
It   is  important   to  appreciate   at  this   point   the  inherent
non-perturbative nature  of the above  formulation, manifesting itself
through the  exponentiation of the $D$  operator. Perturbative results
(at arbitrary  order) may  be recovered as  a special case  through an
appropriate  order-by-order  expansion of  the  above master  formula,
In doing so, one should notice that 
not only $\Gamma_{\phi^n}$ should be expanded starting
from  tree  level, but,  accordingly,  also  the shifts  $\delta\phi$,
$\delta\lambda$  and  $\delta  m^2$,  together with  the  displacement
operator $D$

Since the shifts are treated as independent parameters, the
displacement operator at different perturbative orders commute, and
one can use the ordinary Taylor expansion for the exponentiation of
$D$; up to three loops, one then gets
\begin{equation}
e^D\ =\ 1\: +\: D^{(1)}\: +\: \Big( D^{(2)} +
\frac12D^{(1)\,2}\Big)\: +\: \Big(
D^{(3)}+D^{(2)}D^{(1)}+\frac16D^{(1)\,3}\Big)\: +\: \dots,
\end{equation}
with
\begin{equation}
D^{(n)}\ =\ \delta\phi^{(n)}\frac{\partial}{\partial\phi_{\mathrm R}} +
\delta\lambda^{(n)}\frac{\partial}{\partial\lambdar} + 
\delta m^{2\,(n)}\frac{\partial}{\partial\mr^2}\ .
\end{equation}
The  parameter  shifts   $\delta\phi^{(n)}$, $\delta\lambda^{(n)}$, and
$\delta m^{2\,(n)}$ are loop-wise defined as follows:
\begin{equation}
\delta \phi^{(n)}\ =\ Z^{\frac12\, (n)}_\phi\, \phi_{\rm R}\,,\qquad
\delta\lambda^{(n)}\ =\ Z^{(n)}_\lambda\, \lambda_{\rm R}\,,\qquad
\delta m^{2\,(n)}\ =\ Z^{(n)}_{m^2}\, m^2_{\rm R}\; .
\end{equation}
By acting  with the  operator $e^D$ on  the 1PI  correlation functions
$\phi^n_{\mathrm{R}}\Gamma_{\phi^n}(\lambdar,\mr^2;\mu,\epsilon)$,   we
can easily determine the  expressions for the renormalized correlation
functions $\phi^n_{\mathrm{R}}
\Gamma_{\phi^n}^{\mathrm{R}}(\lambdar,\mr^2;\mu)$  at  the one-,  
and two-loop level, which read
\begin{eqnarray}
  \label{exp} \phi^n_{\mathrm{R}}
\Gamma_{\phi^n}^{\mathrm{R}(1)}(\lambdar,\mr^2;\mu)&=&
\Big<\,D^{(1)}\phi^n_{\mathrm{R}}\Gamma_{\phi^n}^{(0)}(\lambdar,\mr^2;\mu)\,
\Big>\: +\:
\phi^n_{\mathrm{R}}\Gamma_{\phi^n}^{(1)}(\lambdar,\mr^2;\mu,\epsilon)\;,
\nonumber \\[3mm] \phi^n_{\mathrm{R}}
\Gamma_{\phi^n}^{\mathrm{R}(2)}(\lambdar,\mr^2;\mu)&=& \Big<\,\Big(
D^{(2)}+\frac12D^{(1)\,2} \Big)\,\phi^n_{\mathrm{R}}
\Gamma_{\phi^n}^{(0)}(\lambdar,\mr^2;\mu)\: +\: D^{(1)}
\phi^n_{\mathrm{R}}\Gamma_{\phi^n}^{(1)}(\lambdar,\mr^2;\mu,\epsilon)\,\Big>
\nonumber\\
&&+\,\phi^n_{\mathrm{R}}\Gamma_{\phi^n}^{(2)}(\lambdar,\mr^2;\mu,\epsilon)\;.
\end{eqnarray}
The above identities furnish the exact expressions for the 
various renormalization constants, to any given order. 
Specifically, the renormalization constants appearing in the 
$D$-operator [viz.(\ref{Ddef})] are determined through a system of algebraic
equation; the latter is obtained by considering all divergent 
 Green's function $\Gamma_{\phi^n}$ of the theory, and demanding that
the right-hand side of (\ref{exp}) be free of cutoff-dependent terms 
(i.e. no $1/\epsilon$ terms in dimensional regularization).

Let us now list some of the main characteristics and 
advantages of the $D$-formalism,
together with a new, highly non-trivial result obtained through it,  
 and some possible future applications.

(i) The $D$-formalism allows one to to determine unambiguously the CTs
to  any given  order in  perturbation theory:  they  are automatically
obtained through  the straightforward application of  the $D$ operator
on the unrenormalized Green functions, without having to resort to any
additional    arguments    whatsoever.     The   usual    diagrammatic
representation of the CTs is reproduced exactly, through the action of
the $D$  operator on  the Feynman graphs  determining the  given Green
function {\it before} the integration over the virtual loop momenta is
carried  out.  This  is  clearly  an  advantage,  at  least  from  the
logistical point of view,  because it reduces significantly the number
of  Feynman graphs  that  need be  considered  at each  order.  If  one
instead  acts  with  $D$  after  the  momentum  integration  has  been
performed  one  loses  this  direct diagrammatic  interpretation,  but
recovers the same final answer for the renormalized Green function.

(ii) The above  features are particularly relevant in  the context of
gauge  theories,  or  special   gauge  fixing  schemes  (such  as  the
background field  method~\cite{Abbott:1980hw}) where keeping  track of
the  CTs  related  to   the  gauge-fixing  parameter  ($\xi$)  may  be
conceptually subtle.   All such CT are accounted  for unambiguously by
introducing   into   the   $D$   operator   a   term   of   the   form
$\delta\xi\frac\partial{\partial\xi_{\mathrm{R}}}$,  and  subsequently
acting on the corresponding Green's function.

(iii) When  using (\ref{exp}) note  that one must only  include Feynman
diagrams {\it without} counterterms,  i.e.  only those contributing to
the bare  Green's functions. The corresponding  diagrams containing CT
will be  generated {\it dynamically}  by the subsequent action  of the
$D$  operator. The  fact that  the procedure  generates  precisely all
necessary  CT  is reflected  in  the  fact  that the  resulting  final
expression are free of {\it overlapping divergences}.

(iv) The control  that the $D$-formalism provides on  the structure and
organization  of the  CTs to  {\it  all orders}  in perturbation,  has
allowed us to obtain the  {\it exact form} of the deformations induced
due  to the  renormalization procedure  to  any type  of relations  or
constraints  which are  valid  at the  level  of unrenormalized  Green
functions.   In    particular,   as    was   shown   in    detail   in
\cite{Binosi:2005yk},   the   straightforward   application   of   the
$D$-formalism  yields, for  the  first time,  the  deformation of  the
Nielsen   Identities~\cite{Nielsen:1975ph}  in   a  closed,   and,  in
principle,  calculable  form.  This  new result  furnishes  the  exact
dependence  of the  renormalized Green  functions on  the renormalized
gauge-fixing parameter to all orders.
  
(v) 
This formulation opens novel perspectives
for  the  study  of  several  other known  topics.  Specifically,  the
$D$-formalism   may  be   used  to   systematically   investigate  the
renormalization-scheme  dependence of  correlation  functions.  It
may also be employed to algebraically determine the restoring terms of
a ``bad'' UV  regularizing scheme, {\it i.e.}, a  scheme that does not
preserve the Slavnov--Taylor  identities \cite{Slavnov:1972fg}.  
Since it provides all-order
information on the renormalization  of Green functions under study, it
might  be useful  in controlling  the calculation  of non-perturbative
effects,  such  as  those  related  to  the  dynamics  of  renormalons.
The $D$-formalism can be straightforwardly extended to systematize the
procedure  of  renormalizing  non-renormalizable field  theories.   In
particular,  it may be  used to  organize the  infinite series  of CTs
needed  to  renormalize  such  theories.   Even  in  the  case  of
renormalizable  perturbative field theories  the $D$-formalism  can be
automated, for  example with  the aid of  a computational  package, to
reliably compute all  the CTs required for the  renormalization of 1PI
correlation  functions   at  high   orders.

\subsection*{Acknowledgments}
This work was supported by the MCyT grant FPA2002-00612

\end{document}